# Time-resolved vibrational-pump visible-probe spectroscopy for thermal conductivity measurement of metal-halide perovskites

*Shunran Li[1,2], Zhenghong Dai[3], Linda Li[1], Nitin P. Padture[3], Peijun Guo[1,2,*]*

[1]Department of Chemical and Environmental Engineering, Yale University, 9 Hillhouse Avenue, New Haven, CT 06520, USA

[2]Energy Sciences Institute, Yale University, 810 West Campus Drive, West Haven, CT 06516, USA

[3]School of Engineering, Brown University, Providence, RI 02912, USA

Correspondence author: Peijun Guo (peijun.guo@yale.edu)

**Abstract**

Understanding thermal transport at the micro- to nanoscale is crucially important for a wide range of technologies ranging from device thermal management and protection systems to thermal-energy regulation and harvesting. In the past decades, non-contact optical methods such as time-domain and frequency-domain thermoreflectance (TDTR and FDTR) have emerged as extremely powerful and versatile thermal metrological techniques for the measurement of material thermal conductivities. Here, we report the measurement of thermal conductivity of thin films of $CH_3NH_3PbI_3$ ($MAPbI_3$), a prototypical metal-halide perovskite, by developing a time-resolved optical technique called vibrational-pump visible-probe (VPVP) spectroscopy. The VPVP technique relies on the direct thermal excitation of $MAPbI_3$ by femtosecond (fs) mid-infrared (MIR) optical pump pulses that are wavelength-tuned to a vibrational mode of the material, after which the time dependent optical transmittance across the visible range is probed in the ns to the µs time window using a broadband pulsed laser. Using the VPVP method, we determine the thermal conductivities of $MAPbI_3$ thin films deposited on different substrates. The transducer-free VPVP method reported here is expected to permit spectrally resolving and spatiotemporally imaging the dynamic lattice temperature variations in organic, polymeric, and hybrid organic-inorganic semiconductors.

# I. INTRODUCTION

Owing to the developments of experimental and theoretical techniques,[1-12] the understanding of micro- to nanoscale thermal transport in solids has significantly advanced in the past few decades.[13, 14] For the experimental determination of thermal conductivities for solids at the nanoscale (*i.e.*, sub-micron length scale in at least one dimension), time-domain thermoreflectance (TDTR) and frequency-domain thermoreflectance (FDTR) have emerged as popular metrology methods due to their versatility, high accuracy and non-contact nature. Both TDTR and FDTR are pump-probe based techniques, for which a metallic transducer is employed to act as a heat source and simultaneously as a temperature reporter. The interfacial thermal conductance between the transducer and the underlying material under study is usually an independent fitting parameter in the thermal transport model.[1]

In principle, the transducer can be spared if the material of interest can be directly heated by a laser pulse and its optical property provides sensitivity to its lattice temperature. Such transducer-free approach has been pursued in the past for cases of several inorganic semiconductors,[15, 16] where above-bandgap optical pump pulse excites charge carriers from the valence to the conduction band to indirectly heat up the lattice *via* electron-phonon interactions. However, in these transducer-free measurements the charge-carrier relaxation times should be shorter than the characteristic timescales of thermal transport since the excited charge carriers can lead to strong transient optical changes and obscure the lattice temperature evolution. Here, we report a transducer-free optical technique, called vibrational-pump visible-probe (VPVP) spectroscopy, in which we use MIR laser pulses to directly excite material's vibrations to increase the lattice temperature in an impulsive fashion, without the need to excite charge carriers. The subsequent thermal dissipation of the material is monitored by a time-delayed visible probe in the ns to hundreds of μs range that is commensurate with the characteristic timescale of the thermal transport process.

## II. SAMPLE FABRICATION

The benchmark material we adopt in this study is a prototypical metal-halide perovskite (MHP), $CH_3NH_3PbI_3$ (MAPbI$_3$). MHPs are a competitive class of semiconducting materials for cost-effective photovoltaic and light-emitting applications.[17-21] Over the past decade, the power conversion efficiency of MHP-based solar cells has rapidly increased to 25.5%. What's remarkable about MHPs are their solution processability for scalable manufacturing at potentially low cost, as well as their tunable material dimensionalities (3D, 2D, and 1D) and compositions to access diverse properties, enabling promising applications beyond photovoltaics and optoelctronics.[22-24] However, MHPs are thermal 'insulators.' The poor thermal conductivities of MHPs (typically below 1 $W \cdot m^{-1} \cdot K^{-1}$), and of hybrid and organic materials in general,[25-27] mainly stem from the low acoustic phonon velocities[28] and short phonon mean-free paths,[29] and possibly further contributed by the dynamically disordered organic entities in MHPs that lead to phonon scattering.[27, 30] In light of the poor thermal transport and the lack of superior chemical stability in comparison to inorganic semiconductors, fundamental understanding of thermal dissipation and transport processes in MHPs is instrumental for realizing solar cells and other optoelectronic devices with lifetimes on par with their inorganic counterparts, considering the stability of state-of-the-art MHP solar cells is currently in the realm of thousands of hours under realistic operating conditions.[31]

The MAPbI$_3$ films were fabricated according to a literature report.[32] The MAPbI$_3$ precursor solution was prepared by dissolving 159 mg of methylammonium iodide (MAI; Greatcell Solar, Australia), 461 mg of Pb iodide (PbI$_2$; 99.99%, TCI) and 78 mg of dimethyl sulfoxide (DMSO; 99.7%, Acros Organics, USA) in 500 mg of $N,N$-dimethylformamide (DMF; 99.8%, Acros Organics, USA). The substrates were treated with UV-ozone for 15 min to ensure good solution wettability. The MAPbI$_3$ layer was deposited by spin-coating the as-prepared precursor solution at 4000 rpm for 30s with an acceleration of 1300 rpm·s$^{-1}$. At the 10$^{th}$ second of spinning, 250 μL of anti-solvent diethyl ether (DE; 99.7%, Sigma-Aldrich) was dripped at the center. Subsequently, the as-deposited films were annealed at 100 °C for 20 min. This fabrication method consistently yields MAPbI$_3$ films with thickness of ~600 nm. A high-resolution scanning electron microscope (SEM; Quattro ESEM, ThermoFisher Scientific, USA) was used to obtain cross-sectional images of the samples as well as to quantify the film thickness. For static transmission measurements, a deuterium-halogen light source (AvaLight-DHc-S, Avantes) and a fiber-coupled compact spectrometer (AvaSpec-ULS2048CL-EVO, Avantes) were used. Cryogenic measurements were enabled by a liquid-nitrogen optical cryostat (Janis VPF-100) at a vacuum level better than 10$^{-4}$ Torr.

## III. PUMP-PROBE EXPERIMENTAL DESIGN

Figure 1 summarizes the primary optical and electronic components comprising the VPVP setup. A digital delay generator (Hamamatsu C10647; Delay Generator One) produces two synced square-wave electrical pulses at a 1.5-kHz repetition rate; the first pulse triggers another digital delay generator (SRS DG645; Delay Generator Two) and the second pulse acts as the external trigger of the Pharos amplifier output (1.5 mJ pulse energy, 170 fs pulse width, 1030 nm pulse wavelength). Among the 1.5 mJ pulse energy, 0.9 mJ is beam-split into a broadband, high energy MIR optical parametric amplifier (Orpheus-One-HE) to provide the MIR pump pulses with tunable wavelengths. The SRS DG645 outputs two separate, time-delayed square-wave electric pulses, providing the clocks for the optical spectrometer (AvaSpec-ULS2048CL-EVO, 1.5 kHz) and the fiber-coupled supercontinuum laser with ns pulse width (NKT Compact, 1.5 kHz). The supercontinuum laser has a broadband spectral coverage of 450 nm to 2400 nm serving as the probe laser. In each VPVP measurement, the delay time of the probe with respect to the pump is varied from -100 ns to about 100 µs by changing the timing of the trigger signal for the probe laser with the DG645 (Delay Generator Two). Longer delay time window up to 1 ms and beyond can be made possible by simply reducing the repetition rate of the amplifier, although in this work a time window of 100 µs was found sufficient to resolve the entire thermal dissipation process.

The timing jitter of the NKT Compact laser was found to be on the order of 200 ns in our measurements, and the timing uncertainties of the externally triggered Pharos amplifier can be up to one or two oscillator cycles (each cycle is about 13.2 ns). To mitigate these undesirable timing uncertainties and to obtain higher temporal resolution comparable with the pulse width of the probe laser (1~2 ns), the true arrival times of the two lasers are measured pulse by pulse using a time-to-digital converter (Picoquant MultiHarp 150 P). Specifically, less than 0.1% of the probe output is split and fed into a fast silicon photodiode (Thorlabs DET025A), and less than 0.01% of the amplifier output is fed into a second, identical photodiode; in each pump-probe cycle, the time difference between the two pulses is measured using the time-to-digital converter and stored for data post-processing to yield the differential changes in sample transmittance over a range of delay times. This "measure-and-post-process" scheme effectively removes the intrinsic timing jitter of the two lasers and yield a timing accuracy better than 5 ns (see Fig. S1 in the Supplementary Material). We note that during each interval between successive measurements of different delay times set by the DG546, the probe laser is temporarily switched off. However, since the amplifier pump laser cannot be switched off and on as easily as the probe laser, an electric circuit built from a high-bandwidth PNP field-effect transistor is used to suppress the timing signal output on the photodiode from the amplifier (see Fig. S2 in the Supplementary Material) during these intervals.

## IV. EXPERIMENTAL RESULTS

Fig. 2a presents the measured spectral profiles of the MIR pump between 2.5 µm and 8 µm. The spectral width of the pulses ranges from 100 cm$^{-1}$ to 200 cm$^{-1}$, corresponding to bandwidth-limited pulse temporal durations ranging from 170 fs to 330 fs. The spectral profiles were measured using a MIR spectrograph (Horiba µHR) equipped with a 50 grooves·mm$^{-1}$ ruled grating with 6 µm blaze wavelength and a single-channel mercury cadmium telluride (MCT) detector (Infrared Associates). Assuming a 300-µm-diameter spot size of the pump (obtained by focusing the pump using a CaF$_2$ lens with 250 mm focal length), the measured pump power at the tuning range of the OPA permits the calculation of the pump fluences (Fig. 2b). The lowest pump fluence is about 2 mJ·cm$^{-2}$ obtained at 16 µm, so the VPVP method is limited to materials with vibrational modes at wavelengths shorter than 16 µm or wavenumbers higher than 625 cm$^{-1}$.

The MIR transmittance spectra of the 600-nm thick MAPbI$_3$ films on four different types of substrates are shown in Fig. 3. Those substrates are 1-mm thick SiO$_2$ (quartz; SPI supplies, Fused Silica Slides), 2-mm thick CaF$_2$ (ISP Optics), 0.5-mm $c$-plane single-crystalline Al$_2$O$_3$ (sapphire; University Wafers), and 1.1-mm thick SnO$_2$/ITO/glass (MSE supplies). Note that for the SnO$_2$/ITO/glass, the ITO (indium tin oxide) transparent conducting electrode film acts as a reflector in the MIR range owing to the free carrier response (*i.e.*, ITO behaves as a metal), and its transmittance spectrum is not shown in Fig. 3. The attenuated transmittance centered between 3100 cm$^{-1}$ and 3200 cm$^{-1}$, as seen in Fig. 3, stems from the N-H asymmetric stretching on the organic CH$_3$NH$_3^+$ cations in MAPbI$_3$.[33-35] In all subsequent VPVP experiments, we chose to pump the samples at this vibrational resonance. Since the vibrational absorption here is less than 10%, the optical penetration depth of the MIR pump is significantly longer than the 600-nm film thicknesses and as a result the impulsive thermal excitation can be assumed uniform along the thickness of the films. As reported previously by VPVP spectroscopy with picosecond (ps) to single-digit nanosecond (ns) delay time window,[34] fs impulsive excitation of the organic vibration in MAPbI$_3$ is followed by thermal equilibration between the organic and inorganic sublattices, the timescale of which is temperature dependent and ranges from a few hundred ps up to about one ns. Therefore, in the measured ns to µs delay time window investigated here, the entire phonon system of the material is in mutual thermal equilibrium, corresponding to a temperature higher than that of the substrate; the relevant process is thus heat transport within the MAPbI₃ film and from the MAPbI₃ film to the underlying substrates.

The cross-sectional morphology of the MAPbI₃ film on the SnO₂/ITO/glass substrate is shown in Fig. 4a. The SnO₂/ITO/glass substrate comprises a ~185 nm thick ITO film deposited on glass and ~20 nm thick SnO₂ film on the ITO. This type of substrate is frequently used for perovskite solar cell fabrication where SnO₂ serves as an electron-transporting layer.[36] From the SEM image in Fig. 4a it is seen that the grain sizes of MAPbI₃ are in the few-hundred-nm range and on-average is smaller than the thickness of the film; grain boundaries along the lateral dimensions are evident in Fig. 4a. Such grain boundaries are expected to reduce thermal conduction from the film to the substrate, as discussed below. Additional cross-sectional images for the MAPbI₃ films on other types of substrates are shown in Fig. S3 in the Supplementary Material. Interestingly, the MAPbI₃ film deposited on quartz substrate (Fig. S3a in the Supplementary Material) shows similar morphology as that on SnO₂/ITO/glass, both exhibiting rough surfaces and relatively small grain sizes. In contrast, MAPbI₃ films on the *c*-plane sapphire and CaF₂ substrates are notably smoother. Such variations in the film morphology may arise from different surface termination groups that affect the nucleation and growth of the films, and/or from the difference in substrate thermal conductivities which lead to different true annealing temperatures during the MAPbI₃ film fabrication process.

Figure 4b presents the result from steady-state transmission measurements as a function of sample temperature from 78 K to 290 K. Note that the optical density (OD), instead of the transmittance, is plotted here to provide a higher contrast. The abrupt shift of the optical absorption onset wavelength at around 120 K upon cooling is due to the tetragonal-to-orthorhombic phase transition of MAPbI₃. Note that this phase transition temperature is grain-size dependent,[37, 38] and our value here is lower than the single crystal phase-transition temperature of about 160 K.[39, 40] In both the tetragonal and orthorhombic phases, the absorption onset shifts toward higher energy (*i.e.*, toward shorter wavelength) with an increasing temperature. Importantly, there is a one-to-one correspondence between the lattice temperature and the optical absorption (especially in the optical absorption onset wavelength). As a result, the optical absorption can act as a sensitive proxy of the lattice temperature in the time-resolved VPVP experiments, hence transiently probing the optical absorption near the absorption onset wavelength can distill the lattice temperature as a function of time following the impulsive thermal excitation.

Figure 4c shows the transient $\Delta OD$ spectral map of the MAPbI$_3$ film on quartz substrate measured at 78 K in the orthorhombic phase. Here $\Delta OD$ denotes the transient change in optical density, $\Delta OD = OD(t) - OD(0)$, where $OD(t)$ is the optical density at delay time $t$ following the pump excitation, and $OD(0)$ is the optical density without the pump excitation. The $\Delta OD$ spectra exhibit an asymmetric derivative-like line shape, which is composed of a strong transient bleach on the red side and a weak induced absorption on the blue side of the steady-state absorption peak. Comparing it with the static temperature dependent OD spectra (Fig. 4b) indicates that the observed transient optical response corresponds to a blueshift of the absorption onset and with it a lattice temperature increase. This lattice temperature increase is maximized following the pump excitation (*i.e.* near time zero) and decreases in time as the excess heat is dissipated from the MAPbI$_3$ film into the quartz substrate. Since we found that the percentage absorption of the pump by the MAPbI$_3$ film stays invariant with the incident pump power, the initial lattice temperature rise of the film should be linearly proportional to the incident pump fluence. Figure 4d displays the pump fluence dependent kinetic traces of the differential changes in transmittance, denoted as $\Delta Tr/Tr$, extracted at the wavelength of 736 nm. Here Tr stands for optical transmittance and the conversion between $\Delta Tr/Tr$ in Fig. 4d and $\Delta OD$ in Fig. 4c is $\Delta Tr/Tr = 10^{-\Delta OD} - 1$. We find that the decay of $\Delta Tr/Tr$ is independent of the pump fluence, or equivalently is independent of the initial temperature rise, which is a behavior consistent with thermal conduction described by the Fourier's law. It is observed that most of the temperature decay from MAPbI$_3$ to the quartz substrate is completed within a few μs, although tens of μs is required for a complete thermal dissipation (see Fig. S4 and S5 in the Supplementary Material).

## V. EXTRACTION OF THE THERMAL CONDUCTIVITY

In principle, the transient spectral map of Fig. 4c or the kinetic decay trace in Fig. 4d contains all the necessary information to extract the thermal conductivity ($\kappa$) of MAPbI$_3$ and the interfacial thermal conductance ($G$) between MAPbI$_3$ and the substrate. The pumped spot diameter in the VPVP measurements is several hundred μm and is orders of magnitude larger than the film thickness, hence only one-dimensional (1D) cross-plane thermal transport needs to be considered. The lattice temperature $T$ of the MAPbI$_3$ film is a function of time $t$ after the impulsive excitation and the cross-plane coordinate (denoted as $z$), *i.e.*, $T = T(z,t)$. The initial temperature profile can be determined from the measured absorbed pump energy (and with it the absorption coefficient) and the Beer-Lambert law. Using this initial temperature profile and by assuming an arbitrary value for $\kappa$, we can solve a 1D time-dependent heat transport problem based on the Fourier's law of heat conduction and obtain the information of $T(z,t)$. With $T(z,t)$ we can further determine the time dependent film transmittance at a single wavelength or over a spectral range, by discretizing the film into multiple layers along its thickness and identifying each layer with a single temperature and a corresponding optical absorbance. The true values for $\kappa$ and $G$ can then be obtained by sweeping these parameters in the calculations until the predicted time-dependent transmittance matches with the experimental counterpart.

In our analysis of the VPVP results, we make the following simplifying assumptions: (1) the entire film has the same initial temperature and (2) the change in optical transmittance of the film scales linearly with the MAPbI$_3$ temperature rise averaged over the film thickness. Assumption (1) is validated by the small (<10%) vibrational absorption by the film (Fig. 3). Assumption (2) is based on the observed linear relationship (discussed below) between the sample transmittance and the lattice temperature (Fig. 5a); however, it is essentially a Beer-Lambert-level treatment and ignores possible changes in transmittance induced by a gradient in the real part of the optical permittivity. Related to assumption (2), Fig. 5a shows the local (*i.e.*, 80~120 K) variation of steady-state film transmittance with temperature extracted at the peak absorbance wavelength of 78 K (732 nm; see Fig. 4b). Importantly, the transmittance changes linearly with temperature, as long as the temperature is varied locally over small to intermediate range (a few to a couple tens of K) which is satisfied in the VPVP experiments.

Assuming the measured transient change in transmittance at any delay time scales linearly with the thickness-averaged transient temperature rise of the film, the kinetic traces of $\Delta \mathrm{Tr}/\mathrm{Tr}$ (which we choose at the wavelength that shows the strongest signal occurring at ~765 nm), as those shown in Fig. 4d, can be directly converted into the transient decays of the thickness-averaged lattice temperature rise denoted as $\Delta T$. As shown in Fig. 5a, the temperature dependence of the transmittance can be fitted by a linear function $\mathrm{Tr} = AT + B$, or $T = a\mathrm{Tr} + b$, where $\mathrm{Tr}$ is the transmittance and $T$ is the temperature. The differential change in lattice temperature (thickness-averaged) can then be determined as $\Delta T = T - T_0 = (a\mathrm{Tr} + b) - (a\mathrm{Tr}_0 + b) =$ $a(\text{Tr} - \text{Tr}_0) = a\frac{(\text{Tr} - \text{Tr}_0)}{\text{Tr}_0} \times \text{Tr}_0 = a \times \Delta\text{Tr}/\text{Tr} \times \text{Tr}_0$. By performing this analytical procedure using the $\text{Tr}_0$ from steady-state transmittance measurements and the $\Delta\text{Tr}/\text{Tr}$ from VPVP experiments, we can, for instance, obtain the decay of thickness-averaged temperature rise for MAPbI$_3$ on quartz measured at various temperatures from 80 K to 295 K, shown in Fig. 5b. The initial temperature rise is determined to range from about 5 K to 11 K, although higher temperature rise can be easily achieved using a higher pump fluence available in our VPVP setup.

We now focus on the measurement of room-temperature thermal conductivities of MAPbI$_3$ films deposited on different substrates. Note that a determination of the temperature dependent thermal conductivities of MAPbI$_3$ requires a knowledge of the thermal properties of the substrate as a function of temperature, which is outside the scope of the work. By measuring the temperature dependent thermal conductivities of the substrates used in this work by methods such as TDTR, the temperature dependent decay kinetics shown in Fig. 5b can be used to extract the thermal conductivity of MAPbI$_3$ at low temperatures. The transient $\Delta$OD spectral map of MAPbI$_3$ on ITO/SnO$_2$/glass substrate is presented in Fig. 5c. A transient bleach centered at 760~770 nm is observed, a feature arising from lattice temperature rise which is qualitatively similar to the result obtained at 78 K for MAPbI$_3$ on quartz substrate (Fig. 4c). The corresponding normalized lattice temperature decay, $\Delta T$, is plotted in Fig. 5d, which shows that thermal dissipation from the MAPbI$_3$ film to the ITO/SnO$_2$/glass takes more than several µs. This slow process can arise from the low thermal conductivity of the glass substrate, the multiple interfaces involved in thermal transport, and the intrinsically low $\kappa$ of MAPbI$_3$. To corroborate the experimentally inferred lattice temperature decay profile, we performed finite-element calculations using the COMSOL Multiphysics Heat Transfer Module by solving the time-dependent Fourier's law of thermal conduction. In the calculations, the heat capacity (670 J·kg$^{-1}$·K$^{-1}$) and thermal conductivity (1.4 W·m$^{-1}$·K$^{-1}$) of quartz and glass, and the thermal conductivity (10.3 W·m$^{-1}$·K$^{-1}$) of CaF$_2$ were acquired from literature reports.[41-43] The heat capacity of CaF$_2$ (854 J·kg$^{-1}$·K$^{-1}$) and *c*-plane sapphire (770 J·kg$^{-1}$·K$^{-1}$), and the thermal conductivity of *c*-plane sapphire (34 W·m$^{-1}$·K$^{-1}$) were taken from semiconductor manufacturer websites. The density and heat capacity of MAPbI$_3$ are taken from the literature.[28, 40] As demonstrated in Fig. 5d, by using a test value of 0.35 W·m$^{-1}$·K$^{-1}$ for the $\kappa$ of the MAPbI$_3$ film and varying the interfacial thermal conductance $G$ between MAPbI$_3$ and the underlying substrate, a series of calculated decay profiles of the average temperature rise of the MAPbI$_3$ film can be obtained and compared with the experimentally deduced counterpart.

To determine the true $\kappa$ based on the VPVP experimental results, we performed finite-element calculations by sweeping the values of $\kappa$ and $G$ over large respective ranges. This permits us to compute the integrated error by comparing the experimental and calculated results. Specifically, the integrated error is taken as the absolute difference between the computed and experimentally-determined, thickness-averaged temperature rises integrated over the first 5 µs delay time window, $i.e.$ Integrated error = $\int_{t=0\,\mu s}^{t=5\,\mu s} \left| \Delta T_{exp}(t) - \Delta T_{sim}(t) \right| dt$. We found that the integrated error obtained by integration over time windows longer than 5 µs yield identical results since the majority of the signal decays occur during the first 5 µs (Fig. S4 and S5 in the Supplementary Material). Fig. 6a presents the color-coded map of the integrated error as a function of both $\kappa$ and $G$ used in the calculations, for MAPbI$_3$ film on sapphire substrate. We find that the optimal range of $\kappa$, identified from the valleys in the map where the integrated error is minimized, becomes independent of $G$ when $G$ is larger than about $2 \times 10^7$ W $\cdot$ m$^{-2}$ $\cdot$ K$^{-1}$ (which is a range for typical solid-solid interfaces). The insensitivity of $\kappa$ to $G$ arises from the much smaller effective thermal resistance imposed by the interface in comparison to that caused by the 600-nm thick MAPbI$_3$ film alone. For example, a $G = 10^7$ W $\cdot$ m$^{-2}$ $\cdot$ K$^{-1}$ is equivalent to a Kapitza length of $l_K = \kappa/G = 35$ nm (taking $\kappa = 0.35$ W $\cdot$ m$^{-1}$ $\cdot$ K$^{-1}$), which is significantly less than the 600-nm film thickness. The sensitivity of the VPVP approach is reflected by the narrowness of the valley in the map of the integrated error, which is better illustrated in Fig. 6b that plots the integrated error as a function of $\kappa$ for a fixed $G$ value of $8 \times 10^7$ W $\cdot$ m$^{-2}$ $\cdot$ K$^{-1}$ (a value that doesn't matter as long as it is larger than $2 \times 10^7$ W $\cdot$ m$^{-2}$ $\cdot$ K$^{-1}$). By identifying the $\kappa$ that results in the smallest integrated error and proposing an error bar in this approach to be obtained when the $\kappa$ leads to a 20% increase in the integrated error from its minimal value, we can determine the true $\kappa$ of MAPbI$_3$ on sapphire to be $0.41 \pm 0.03$ W $\cdot$ m$^{-2}$ $\cdot$ K$^{-1}$. As shown in Fig. 6c, using this value of $\kappa$ in the finite-element calculation yields a decay curve that matches very well with the experimental counterpart. We note that our measurements here based on thick MAPbI$_3$ films do not provide quantitative information on the values of $G$. More reliable values for $G$ can be further pursued by measuring MAPbI$_3$ films with sub-hundred nm thicknesses. However, thinner films will likely possess different film morphologies and grain size distributions, the investigation of which is beyond the scope of this work.

In Fig. S4 of the Supplementary Material we summarize the results for MAPbI$_3$ films on quartz, ITO/SnO$_2$/glass and CaF$_2$ substrates, respectively. In all three cases, the choice of $G$ does not influence the dependence of the integrated error on $\kappa$ when $G$ exceeds $2 \times 10^7$ W·m$^{-2}$·K$^{-1}$, for the same reason as for the case of MAPbI$_3$ on sapphire substrate. We found that the $\kappa$ of MAPbI$_3$ on $c$-plane sapphire (~0.41 W·m$^{-1}$·K$^{-1}$) and CaF$_2$ (~0.51 W·m$^{-1}$·K$^{-1}$) are higher than those on quartz and ITO/SnO$_2$/glass substrates. This trend of thermal conductivity variation with the substrate type is consistent with the substrate-dependent thin film morphologies from the SEM images. Overall, the measured values here are comparable to those obtained previously using other methods,[28, 44-47] including TDTR and FDTR, and note that a substrate dependent thermal conductivity has been observed previously.[48] We note that $\kappa$ for the MAPbI$_3$ on $c$-plane sapphire and CaF$_2$ lie between polycrystalline powders and single-crystalline MAPbI$_3$, suggesting that film morphological control can effectively boost the thermal conductivity of MAPbI$_3$, which can be beneficial for the thermal management of MHP-based devices.

## VI. CONCLUSION

In conclusion, we demonstrate VPVP as a viable optical metrological tool to characterize the thermal transport properties of prototypical MPH, MAPbI$_3$. Compared to the well-established thermoreflectance techniques, the VPVP method does not require a metallic transducer and hence do not involve an interfacial thermal conductance between the perovskite and the metallic transducer. Importantly, the VPVP provides direct information on the local temperature, which opens a possibility of spatiotemporal thermal imaging at both near-equilibrium and far-from-equilibrium conditions, the latter made possible by direct impulsive thermal stimulation of the targeted material with higher-energy MIR pulses than used in this work. We anticipate that the demonstrated VPVP technique can be employed for investigating thermal transport in disordered materials such as organic or polymeric semiconducting films,[49] metal-organic frameworks,[50] covalent organic frameworks,[51] biological materials,[52] and low-dimensional halide perovskites,[53-56] some of which may not be amenable to high-vacuum deposition of metallic transducers atop. We do note that due to the spectral coverage of commercial optical parametric amplifiers (OPAs), the proposed method does not readily apply to inorganic materials that lack vibrational modes shorter than 16 µm (*i.e.*, 625 cm⁻¹), although far-infrared OPAs and intense terahertz source are being actively developed for coherent phonon excitations of solids with vibrational modes in the corresponding spectral ranges.[57, 58]

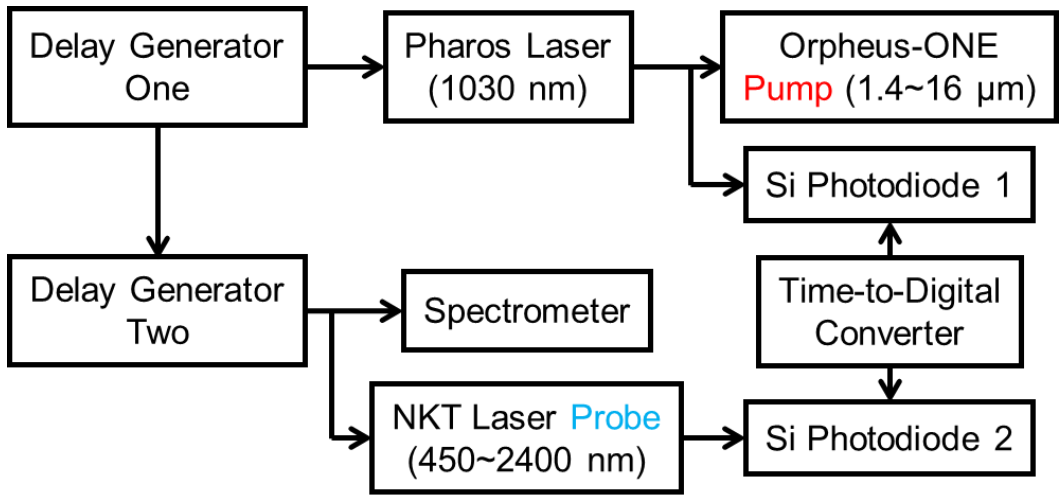

**Figure 1**. Schematic diagram of the VPVP spectroscopic setup showing the key optical and electronic components.

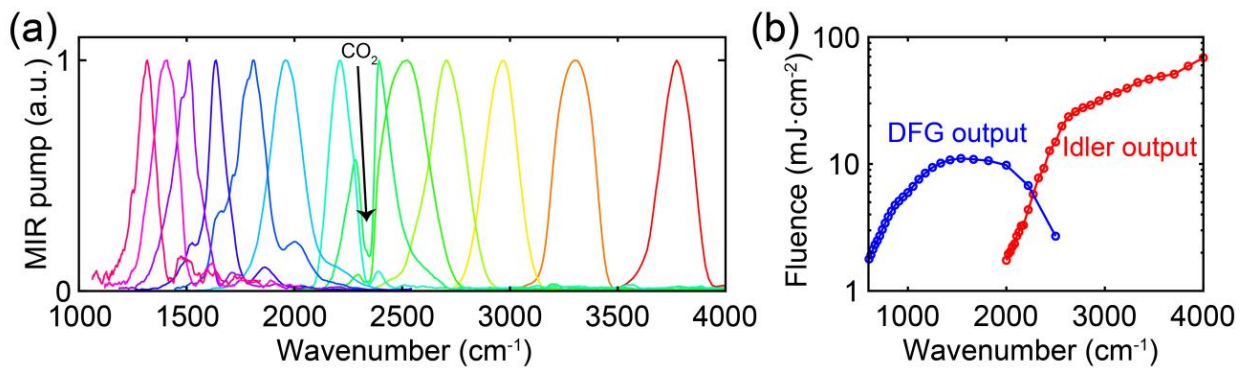

**Figure 2**. (a) Measured spectra of the infrared pump output by the MIR OPA (Orpheus-One-HE). The ambient $CO_2$ absorption peak is intentionally included to demonstrate the pump spectral bandwidth. (b) Calculated pump fluence as a function of wavenumber using the measured output powers and assuming a circular spot with 300 µm diameter.

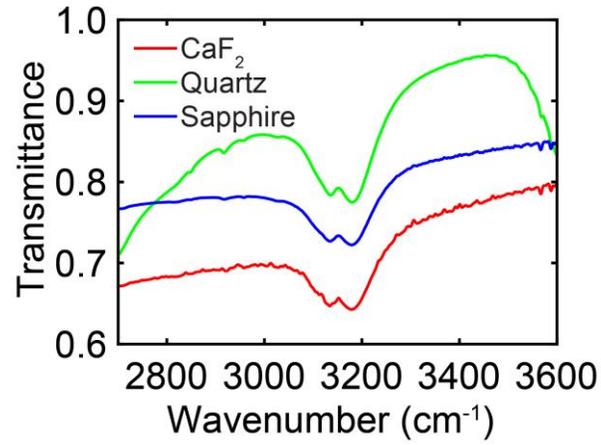

**Figure 3**. MIR transmittance spectra (2700~3600 cm$^{-1}$) of the 600-nm thick MAPbI$_3$ films on various substrates. The curves are offset by 0.1 for clarity.

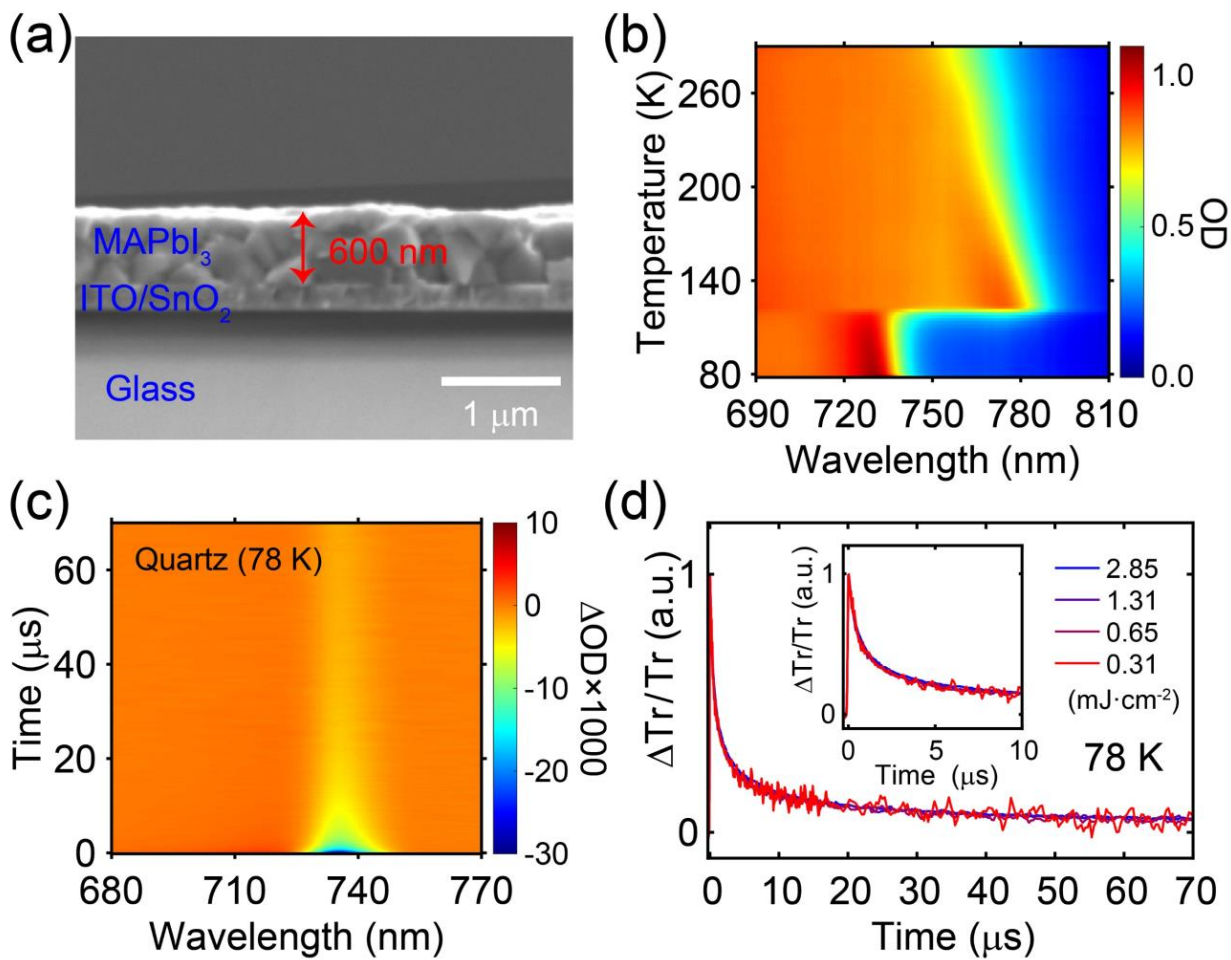

**Figure 4**. (a) Cross-sectional SEM image of the MAPbI$_3$ film deposited on ITO/SnO$_2$/glass substrate. (b) Steady-state, temperature dependent optical density (OD) of the MAPbI$_3$ film on quartz substrate. (c) Transient $\Delta$OD spectral map of the MAPbI$_3$ film on quartz substrate measured at 78 K pumped at 3170 nm with a pump fluence of 2.85 mJ·cm$^{-2}$. (d) Kinetic traces of the differential change in transmittance ($\Delta$Tr/Tr) at 736 nm, plotted over the time window of 0~70 μs for various pump fluences ranging from 0.31 mJ·cm$^{-2}$ to 2.85 mJ·cm$^{-2}$; inset shows the same data over 0~10 μs time window.

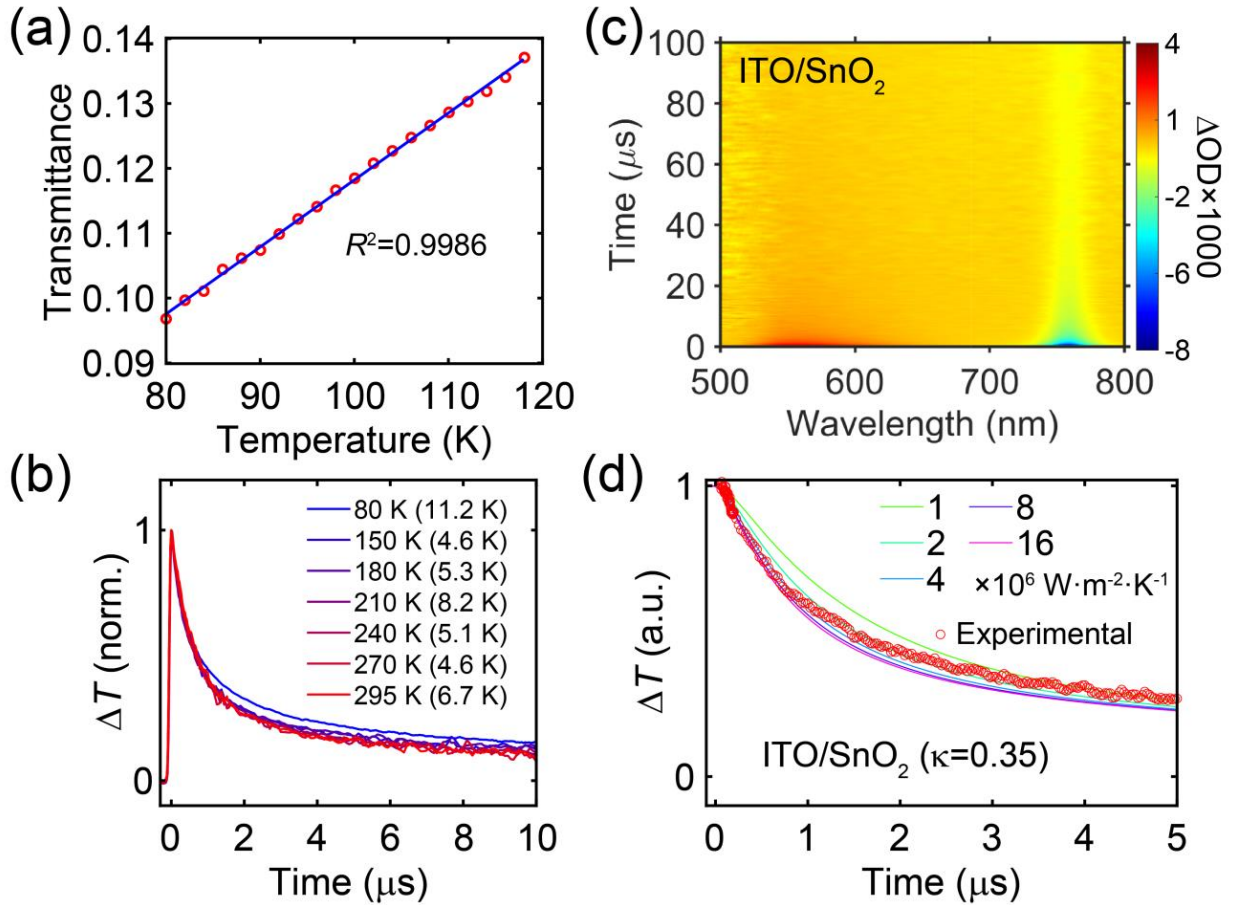

**Figure 5**. (a) Temperature dependence of the steady-state transmittance at the exciton absorption peak (732 nm) for the MAPbI$_3$ film on quartz substrate, showing a linear dependence of the transmittance on the lattice temperature. (b) Extracted temperature decays of the MAPbI$_3$ film on quartz substrate measured at different temperatures ranging from 78 K to 295 K. The maximal temperature rises are indicated in the inset. (c) Transient ΔOD spectral map of the MAPbI$_3$ film on ITO/SnO$_2$/glass substrate measured at room temperature with 3170-nm pump. (d) Comparison of the experimentally deduced MAPbI$_3$ lattice temperature decay (red dotted line) with those calculated using various MAPbI$_3$-SnO$_2$ interfacial thermal conductance (curves); the thermal conductivity of MAPbI$_3$ film is fixed at 0.35 W·m$^{-1}$·K$^{-1}$ in the calculations.

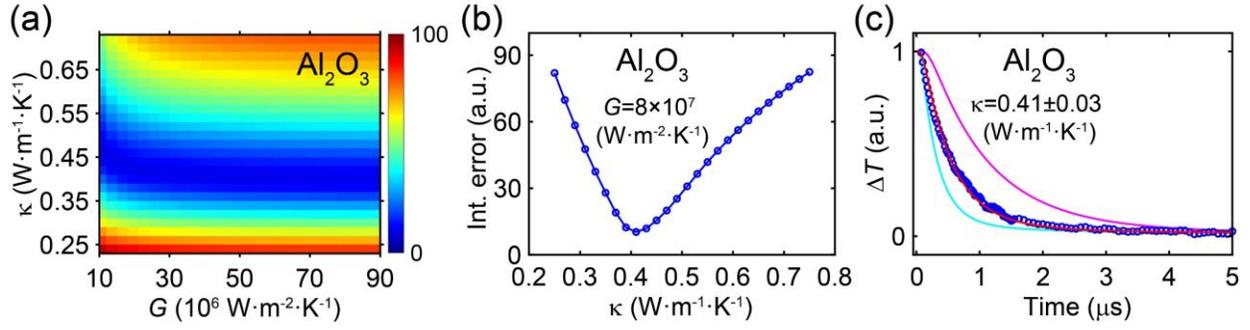

**Figure 6**. (a) Color-coded map of the integrated error (defined in the text) as a function of $\kappa$ and $G$ values. (b) Integrated error as a function of $\kappa$ under a fixed $G$ value of $8 \times 10^7$ W·m⁻²·K⁻¹. (c) Temperature decay profiles from finite-element computation (red curve) and experiments (blue open circles) using the optimal $\kappa$ value of 0.41 W·m⁻¹·K⁻¹ ($G$ value is taken as $8 \times 10^7$ W·m⁻²·K⁻¹). The magenta curve shows computation result using the smallest value for $\kappa$ (0.24 W·m⁻¹·K⁻¹) in the parametric sweeps and the cyan curve shows computation result using the largest values for $\kappa$ (0.69 W·m⁻¹·K⁻¹); $G$ is fixed as $8 \times 10^7$ W·m⁻²·K⁻¹. All the results shown are for the MAPbI₃ film on sapphire substrate.

**Supporting Information**

Additional figures and discussion can be found in the Supporting Information.

**Correspondence**

E-mail: peijun.guo@yale.edu

**Acknowledgements**

S.L., L.L., and P.G. acknowledge Yale University's lab set-up fund. Z.D. and N.P.P. acknowledge the support from the US Office of Naval Research (grant no. N00014-20-1-2574).

**Data Availability**

The data that support the findings of this study are available from the corresponding author upon reasonable request.

# Supplementary Material

## Time-resolved vibrational-pump visible-probe spectroscopy for thermal conductivity measurement of metal-halide perovskites

*Shunran Li[1,2], Zhenghong Dai[3], Linda Li[1], Nitin P. Padture[3], Peijun Guo[1,2,*]*

[1]Department of Chemical and Environmental Engineering, Yale University, 9 Hillhouse Avenue, New Haven, CT 06520, USA

[2]Energy Sciences Institute, Yale University, 810 West Campus Drive, West Haven, CT 06516, USA

[3]School of Engineering, Brown University, Providence, RI 02912, USA

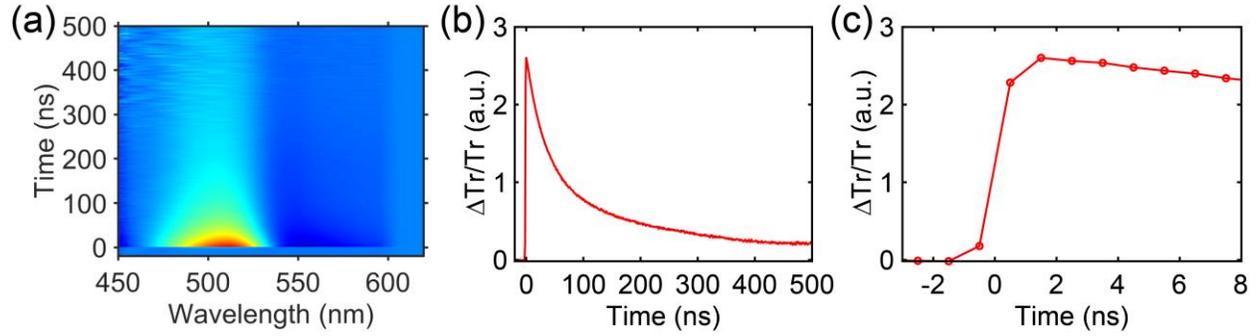

**Figure S1.** Transient $\Delta Tr/Tr$ spectral map at early delay times taken from an indium tin oxide nanorod array pumped by 3000-nm laser pulses. (a) Transient $\Delta Tr/Tr$ spectral map. (b) Kinetic decay trace of the $\Delta Tr/Tr$ at 500 nm from -20 ns to 500 ns. (c) Kinetic decay trace of the $\Delta Tr/Tr$ at 500 nm from -3 ns to 8 ns. The ITO nanorod array has sub-picosecond electron-phonon coupling dynamics, so the true transient response is a step function at time zero.[1, 2]

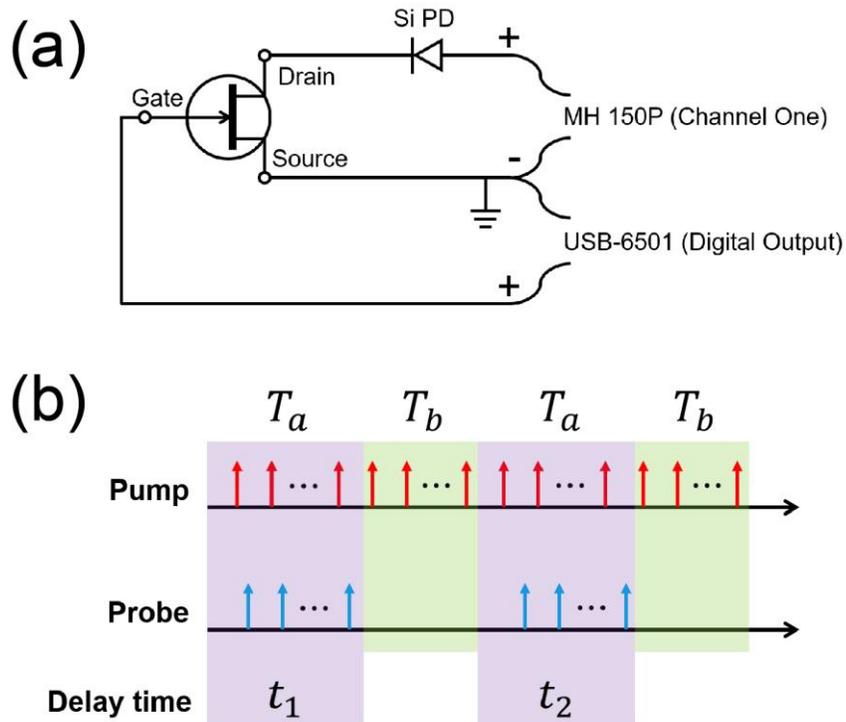

**Figure S2.** (a) Schematic diagram of the timing circuit for measuring the arrival time of the pump pulses (Si PD: silicon photodiode. MH 150P: MultiHarp 150 P Time-to-Digital Converter from Picoquant). The USB-6501 (National Instruments) is used to turn off a PNP transistor at the time intervals $T_b$, during which the probe is off and the DG645 is being configured to change the delay time of the probe. The spectrometer data acquisition is active only during the time intervals $T_a$. (b) Schematic illustration showing the time intervals $T_a$ (probe on; spectrometer on) and $T_b$ (probe off; spectrometer off) of the TA measurements. Note that the pump cannot be turned off during $T_b$, so only the photodiode which receives signal from the pump is being turned off (by the transistor).

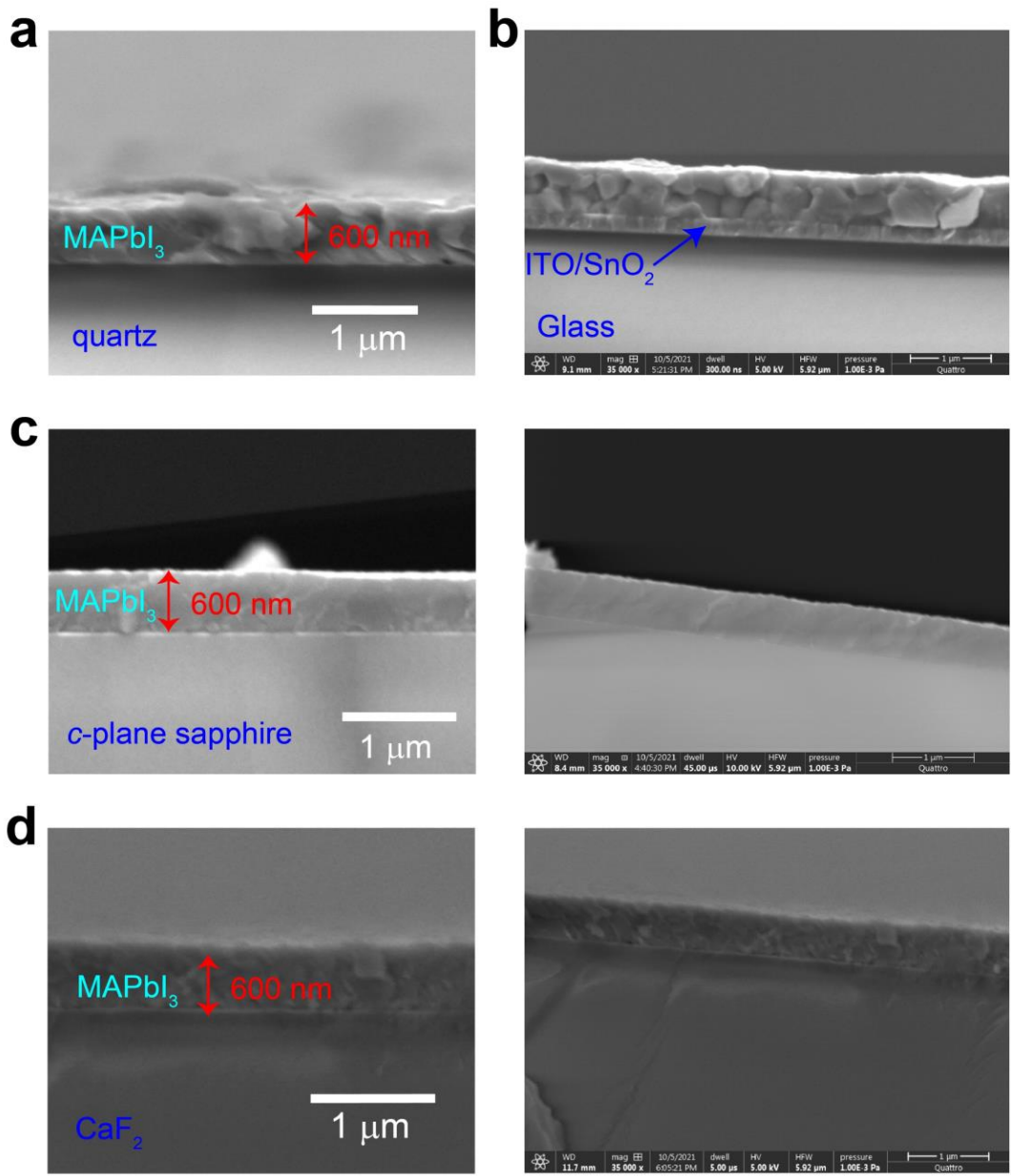

**Figure S3.** Cross-sectional SEM images of MAPbI₃ films on various substrates: (a) quartz, (b) ITO/SiO₂/glass, (c) *c*-plane sapphire, and (d) CaF₂.

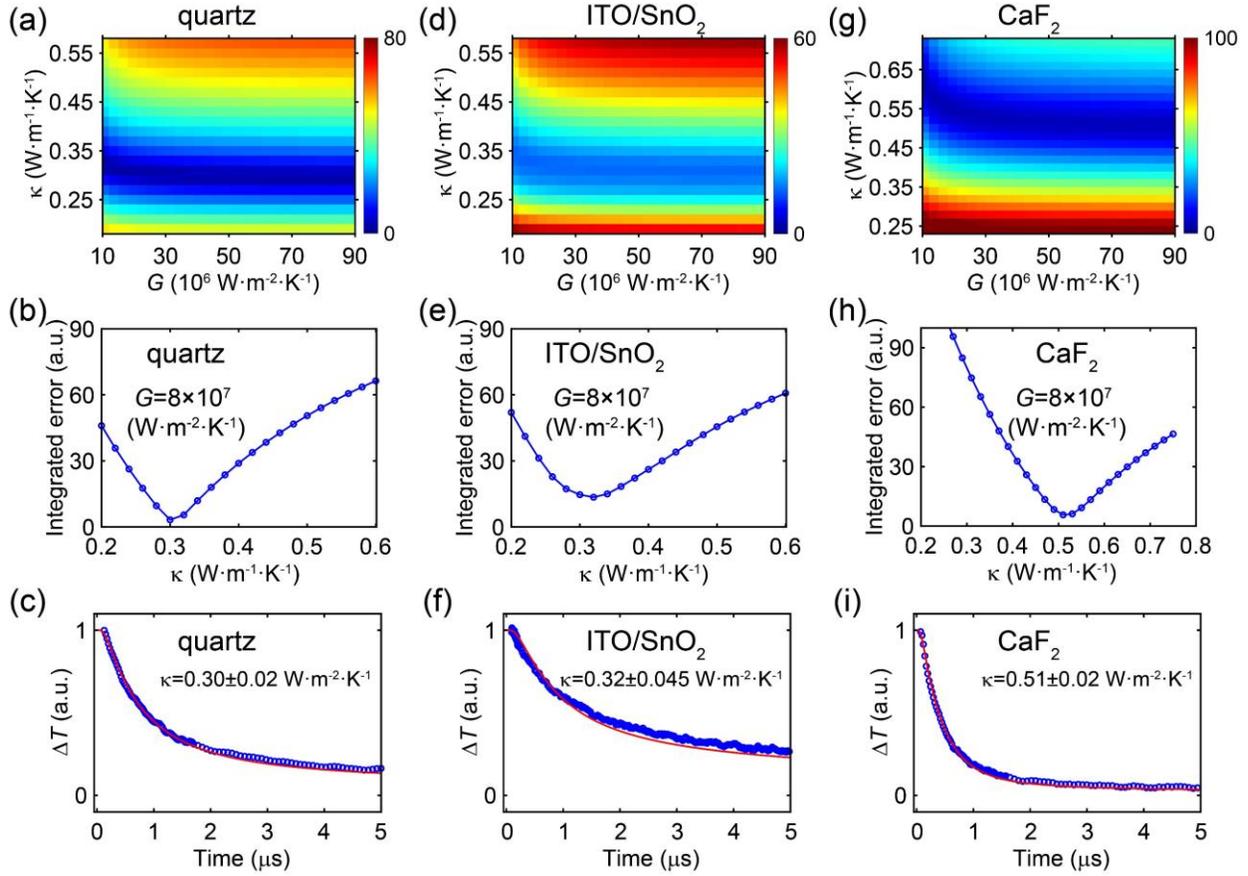

**Figure S4.** Sensitivity analysis by comparing the simulated and experimental temperature decay profiles. The integrated error, shown in (a), (d) and (g) as color-coded maps, are defined as the *absolute* difference between the finite-element calculated and experimentally-determined temperature rises (thickness-averaged) integrated over the first 5 μs delay time window, i.e. Integrated error = $\int_{t=0\ \mu s}^{t=5\ \mu s} |\Delta T_{exp}(t) - \Delta T_{sim}(t)| dt$. The Integrated error is calculated for a wide range of $G$ and $\kappa$ values used in the simulations, where $G$ is the interfacial thermal conductance between the MAPbI₃ film and the substrate, and $\kappa$ is the thermal conductivity of the MAPbI₃ film. The plots in (b), (e) and (h) are the Integrated error v.s. $\kappa$ under a fixed $G$ value of $8\times10^7$ W·m⁻²·K⁻¹. The plots in (c), (f) and (i) are the temperature decay profiles from simulations (red curve) and experiments (blue open circles) using the optimal $\kappa$ values (indicated in the insets). Results shown in (a), (b) and (c) are for the MAPbI₃ film on quartz substrate; results shown in (d), (e) and (f) are for the MAPbI₃ film on ITO/SnO₂/glass substrate; results shown in (g), (h) and (i) are for the MAPbI₃ film on CaF₂ substrate.

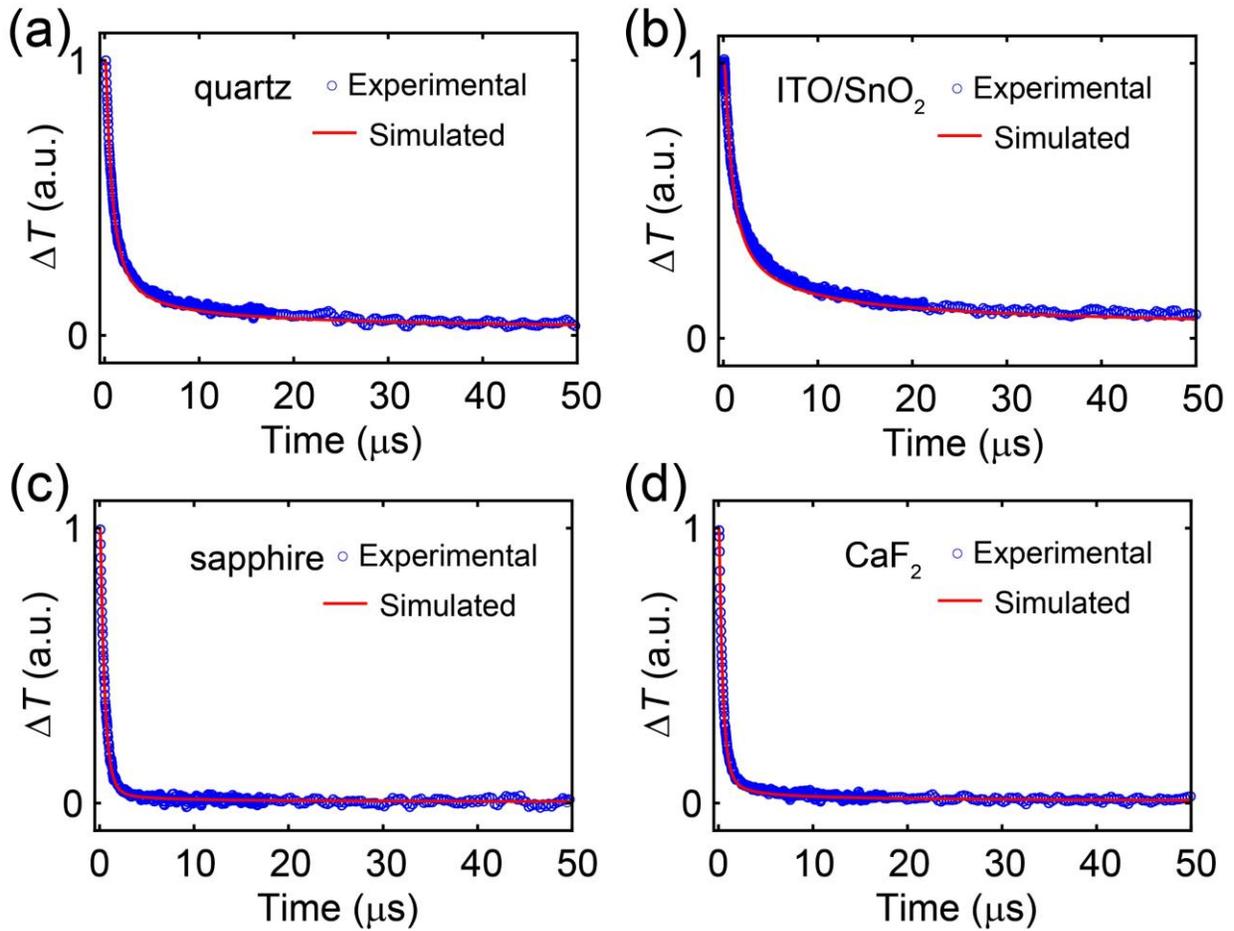

**Figure S5.** Temperature decay profiles from simulations (red curve) and experiments (blue open circles) using the optimal $\kappa$ values, plotted for the time window up to 50 µs. The results shown in (a), (b), (c) and (d) are for quartz, ITO/SnO$_2$/glass, *c*-plane sapphire and CaF$_2$ substrates, respectively.